\documentclass[%
 reprint,
 amsmath,amssymb,
 aps,
pra,
]{revtex4-1}

\usepackage{dsfont}
\usepackage{graphicx}% Include figure files
\usepackage{dcolumn}% Align table columns on decimal point
\usepackage{bm}% bold math
\usepackage{bbm}
\usepackage{soul}
\usepackage{hyperref}
\usepackage{cleveref}
\newcommand{\be}{\begin{equation} }
\newcommand{\ee}{\end{equation} }
\newcommand{\ux}{\underline{\sigma} }
\newcommand{\us}{\underline{s} }

\newcommand{\bsl }{\backslash}

\newcommand{\hi}{\widehat{i} }
\newcommand{\hk}{\widehat{k} }
\newcommand{\hj}{\widehat{j} }
\newcommand{\cut}[1]{{}}
\newcommand{\ra}{\rightarrow}

\DeclareMathOperator\arctanh{arctanh}

\begin{document}

%\preprint{APS/123-QED}
 \title{Slow Spin Dynamics and Self-Sustained Clusters in Sparsely Connected Systems}% Force line breaks with \\
%DS \title{Slow Spin Dynamics and Self-Sustained Clusters in Sparse Graphs}% Force line breaks with \\
%\thanks{A footnote to the article title}%

\author{Jacopo Rocchi}
\email{j.rocchi@aston.ac.uk}
 \affiliation{
 Nonlinearity and Complexity Research Group, Aston University, Birmingham B4 7ET, United Kingdom
}
\author{David Saad}%
 \email{d.saad@aston.ac.uk}
 \affiliation{
 Nonlinearity and Complexity Research Group, Aston University, Birmingham B4 7ET, United Kingdom
}
\author{Chi Ho Yeung}
\email{chyeung@eduhk.hk}
\affiliation{Dept. of Science and Environmental Studies, The Education University of Hong Kong, Hong Kong}
%\collaboration{MUSO Collaboration}%\noaffiliation

\date{\today}% It is always \today, today,
             %  but any date may be explicitly specified

\begin{abstract}

To identify emerging microscopic structures in low temperature spin glasses, we study self-sustained clusters (SSC) in spin models defined on sparse random graphs. A message-passing  algorithm is developed to determine the probability of individual spins to belong to SSC. Results for specific instances, which compare the predicted SSC associations with the dynamical  properties of spins obtained from numerical simulations, show that SSC association identifies individual slow-evolving spins. This insight gives rise to a powerful approach for predicting individual spin \emph{dynamics} from a single snapshot of an equilibrium spin configuration, namely from limited \emph{static} information, which can be used to devise generic prediction tools applicable to a wide range of areas.

%\begin{description}
%\item[Usage]
%Secondary publications and information retrieval purposes.
%\item[PACS numbers]
%May be entered using the \verb+\pacs{#1}+ command.
%\item[Structure]
%You may use the \texttt{description} environment to structure your abstract; use the optional argument of the \verb+\item+ command to give the category of each item.
%\end{description}
\end{abstract}

\pacs{Valid PACS appear here}% PACS, the Physics and Astronomy
                             % Classification Scheme.
%\keywords{Suggested keywords}%Use showkeys class option if keyword
                              %display desired
\maketitle

%\tableofcontents
Spin glass models of disordered systems are characterized by a rich structure of the free-energy landscape and a non-trivial dynamics at low temperatures.
Mean field analyses~\cite{sherrington1975solvable, parisi1979infinite, young1998spin} typically characterize the static properties of models based on a set of macroscopic order parameters~\cite{mezard1990spin, Parisi2007}, providing insight into their equilibrium properties and dynamical behaviour~\cite{bouchaud1998out, marinari2000replica}.

Fully and sparsely connected models have been extensively studied using powerful tools such as the replica and cavity methods~\cite{mezard1990spin,mezard2001bethe}. While the different temperature regimes are well understood in terms of the (free-) energy landscape, it is more difficult to describe their manifestation at the microscopic level.
Interesting cases where this connection is clearer are constraint satisfaction problems studied at zero temperature, which give rise to solution clusters containing frozen variables in intermediate regimes prior to the satisfiability transition~\cite{mezard2003two, krzakala2007gibbs, montanari2008clusters, achlioptas2009random, semerjian2008freezing, braunstein2016large}. In addition to the insight gained, understanding the microscopic properties and their links to the system dynamics are essential for devising approximate optimization algorithms for specific instances. 

Studies of the relation between system equilibrium properties and its dynamical behaviour~\cite{barrat1999time, ricci2000glassy} show that it is possible to interpret system dynamical characteristics in terms of its ground-state structural properties. Moreover, they 
reveal interesting properties of the low temperature dynamics such as the spontaneous time-scale separation between slow and fast evolving spins.
This phenomenon has been observed also in numerical experiments in systems defined on finite dimensional lattices~\cite{roma2006signature, roma2010domain}, giving rise to the notion of \emph{rigidity lattice}~\cite{barahona1982morphology}. 
Another approach linking equilibrium and dynamical properties~\cite{lage2014message}, reveals that metastable states relate to fixed points of the generalised Belief Propagation (BP)  in the 2D Edwards-Anderson model.

Our approach aims to link equilibrium and dynamical properties of spin models on random graphs via the concept of \emph{Self-Sustained Clusters} (SSC) and the utilization of BP methods. The central objects of our approach are SSC, introduced in the study of the SK~\cite{yeung2013self} and 3-spin Ising models~\cite{rocchi2016self}. By definition, the field induced by in-cluster spins, on spins belonging to an SSC, has a higher magnitude than the one induced by out-cluster spins. SSC can be regarded as metastable formations that correspond to suboptimal system configurations and explain the emergence of slow-evolving spins in fully connected models.

Here we focus on systems with simple discrete disorder ($J \!=\! \pm 1$) defined on Erdos-Renyi (ER) and Random Regular (RR) graphs. We develop and apply a BP algorithm to identify SSC variables in these settings and run dynamical simulations at different temperatures to study the relation between slow-evolving spins and SSC variables. 
More precisely, after equilibration for a long time, we sample a configuration and infer its SSC structure; this is then contrasted against its microscopic dynamics, starting from this configuration and monitoring the dynamical properties of individual spins, such as their flipping probability. We identify a strong relation between these two quantities,
which can be used as a powerful tool to predict individual spin dynamics based on a single snapshot of equilibrium spin configuration. After defining the SSC framework of sparse graphs, we give details on the protocol employed for the simulations and analyze the obtained results.

\emph{SSC formation:}
Consider a pairwise Ising model on a sparse graph $\mathcal{G}$ with link weights $\{J_{ij}\}$ and a given spin configuration $\underline{s}$. We introduce an SSC membership variable $\sigma_i\!=\!1$ to indicate that spin $i$ belongs to an SSC and $\sigma_i\!=\!0$ otherwise. This SSC condition is enforced by an indicator function, taking the value of one if the cluster membership condition is obeyed and zero otherwise 
\be
\mathds{I}(\ux|\us,\mathcal{G}) = \prod_{i=1}^{N} \left \{ (1-\sigma_i) +\sigma_i \theta [u_i^2 - v_i^2 ] \frac{}{} \right \}~,
\label{eq:glocon}
\ee
where $u_i$ and $v_i$ are the in-cluster and out-cluster induced fields, respectively, defined by 
\begin{equation}
u_i \!=\! \sum_{m \in \partial i} J_{im} \sigma_m s_m ~\mbox{ and }~
v_i \!=\! \sum_{m \in \partial i} J_{im} (1\!-\!\sigma_m) s_m~,
\end{equation}
%\begin{eqnarray}
%u_i &=& \sum_{m \in \partial i} J_{im} \sigma_m s_m \\
%v_i &=& \sum_{m \in \partial i} J_{im} (1-\sigma_m) s_m~,
%\end{eqnarray}
where $\partial i $ denotes the set of neighbours of node $i$. The sum of the two fields constitutes the total field 
\be
h_i = \sum_{m \in \partial i} J_{im}  s_m = u_i + v_i\:.
\ee
The resilience parameter $\epsilon$ is defined by the condition $u_i^2 > v_i^2 + \epsilon$ and indicates the strength of the cluster.
The set of all the possible $\ux$ vectors that satisfy Eq.~(\ref{eq:glocon}) corresponds to all the possible SSC structures given the spin configuration $\underline{s}$.
It is important to notice that the trivial realisations $\sigma_i=0 \: \forall i$  and $\sigma_i=1 \: \forall i$ satisfy Eq.~(\ref{eq:glocon}): they correspond to the cases where there are no SSC or when the only SSC is the complete system, respectively. 

\begin{figure}[ht]
\centering
\includegraphics[width=85mm]{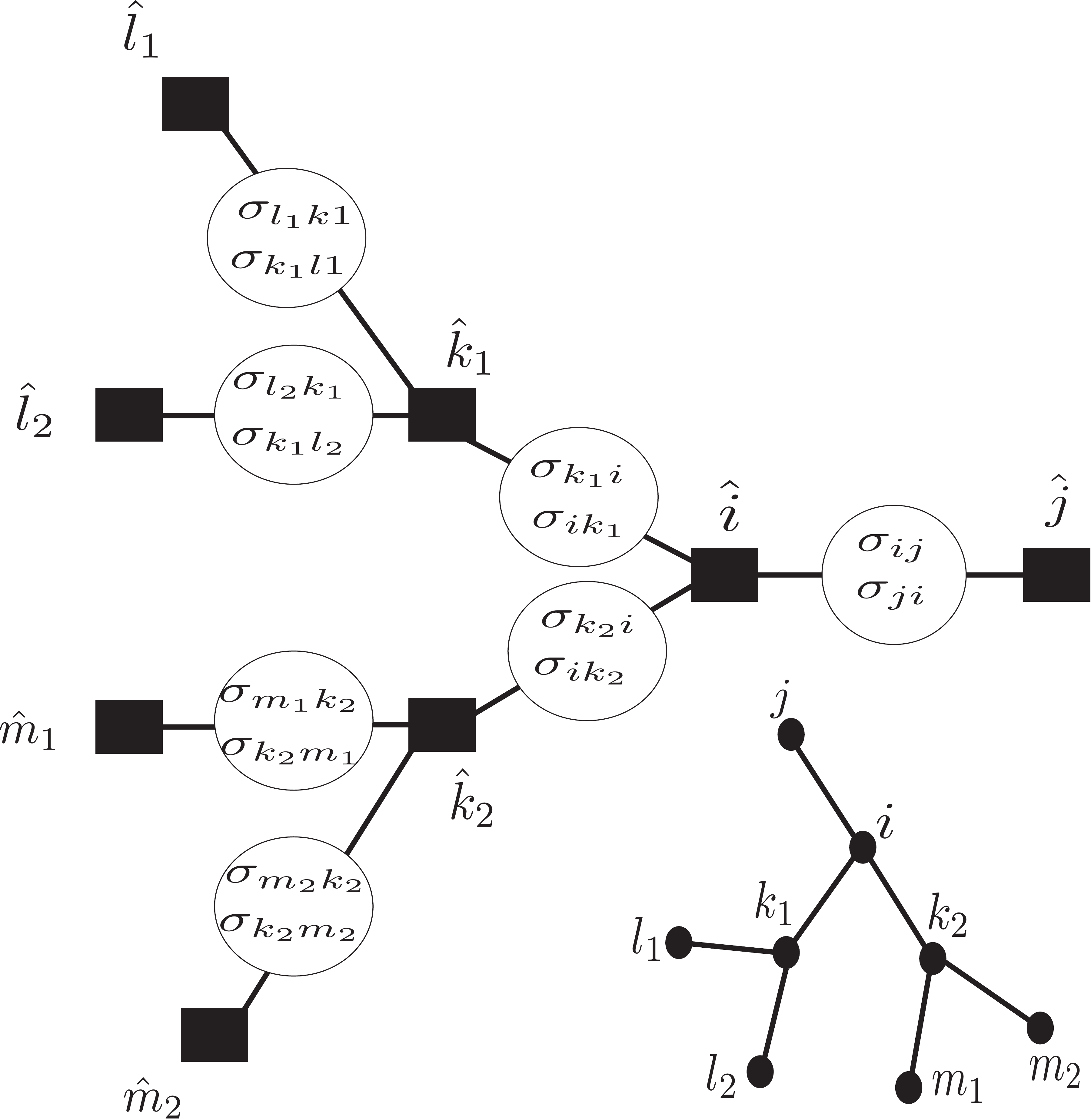}
\caption{ {A super-factor graph corresponding to the original graph at the bottom right of the figure. Each super-variable contains two variables of the original graph. For each node of the original graph $\mathcal{G}$ we have a super-factor and for each link we have a super-variable. All the variables $\sigma_{ik}, \: \forall k \in \partial i$ are forced to be the same by the super-factors' constraint. The super-factor and original graphs share the same topology.} }
\label{fig_Fig1}
\end{figure}

\emph{BP equations:}
The variable $\sigma_i$ depends on the state of the variables {$\sigma_{j \in \partial i}$} defined on all the neighbours of node $i$.
It is convenient to define a factor graph, but a conventional factor graph {gives} rise to many short loops that hamper convergence and proper calculation of the marginals. As in similar cases~\cite{wong2008DiversityColoring}, we will introduce a super-factor graph of super-variables $S_{ij} = (\sigma_{ij},\sigma_{ji})$ comprising the states of variable pairs $ \sigma_{ij} \!=\! \sigma_i $ and $\sigma_{ji} \!=\! \sigma_j$.  {The variable $\sigma_i$ is copied in all the super-variables associated with node $i$, as shown in Fig.~\ref{fig_Fig1}, while the different indices $j$ in $\sigma_{ij}$ refer to neighboring links}. We consider the double-index variables as independent states but enforce their equality through the super-factors, 
\be
\psi_{\hi}^{S} (\underline{S}_{\partial \hi}) =  \psi_{\hi}(\ux_{\partial \hi}) \prod_{k \in \partial i}  \delta_{\sigma_{ik}, \sigma_{i}}\:
\ee
defined on the original node $i$, and denoted by $\hi$ in the super-factor graph, where $\psi_{\hi}(\ux_{\partial \hi})$ is given by
\begin{equation}
\psi_{\hi}(\ux_{\partial \hi}) = \left \{ (1-\sigma_i) + \sigma_i \theta[u_i^2 - v_i^2-\epsilon] \frac{}{} \right \} \:.
\end{equation}
More details are provided in Section~1 of the Supplementary Material (SM). 

The BP-equations for this super-graph read
\be
\nu_{\hi \ra (ij)} (S_{ij}) \propto \sum_{\underline{S} \in \partial \hi \bsl S_{ij}} \psi_{\hi}^{S} (\underline{S}_{\partial \hi}) \prod_{k \in \partial i \bsl j} \eta_{(ki)\ra \hi} (S_{ki})
\label{eq:nuSFG}
\ee
where, given the pairwise nature of the interactions in the original graph, messages from the super-variables to the super-factors take the form
\be
\eta_{(ij) \ra \hj} (S_{ij}) \propto \nu_{\hi \ra (ij)} (S_{ij}).
\label{eq:etaSFG}
\ee
These equations can be simplified in a single line to
\be
m_{i \ra j} (\sigma_i, \sigma_j) \propto \sum_{\underline{\sigma}_{\partial i \bsl j}} \psi_{\hi}(\underline{\sigma}_{\partial \hi})  \prod_{k \in \partial i \bsl j} m_{k \ra i} (\sigma_k, \sigma_i)
\label{mitoj}
\ee
where we define the messages $m_{i \ra j} (\sigma_i, \sigma_j) \!=\! \eta_{(ij) \ra \hj} (S_{ij}) \!=\! \nu_{\hi \ra (ij)} (S_{ij})$. Marginals of the super-variables can be computed  from the equation
\be
\eta_{(ij)} (S_{ij}) \propto \nu_{\hi \ra (ij)} (S_{ij})  \nu_{\hj \ra (ij)} (S_{ji}) 
\ee
and can be used to compute single-node marginals
\be
p(\sigma_i) = \sum_{\sigma_j} \eta_{(ij)} (S_{ij}) = \sum_{\sigma_j} m_{i \ra j} (\sigma_i, \sigma_j) m_{j \ra i} (\sigma_j, \sigma_i)\:.
\ee
Single-node marginals have to satisfy a consistency check since $p(\sigma_i)$ could be computed from $\sum_{\sigma_k} \eta_{(ik)} (S_{ik})$ for all nodes $k \!\in\! \partial i $. Thus, after convergence, this condition is checked to assess the quality of the obtained marginals. 

\begin{figure}[ht]
	\centering
	\includegraphics[width=85mm]{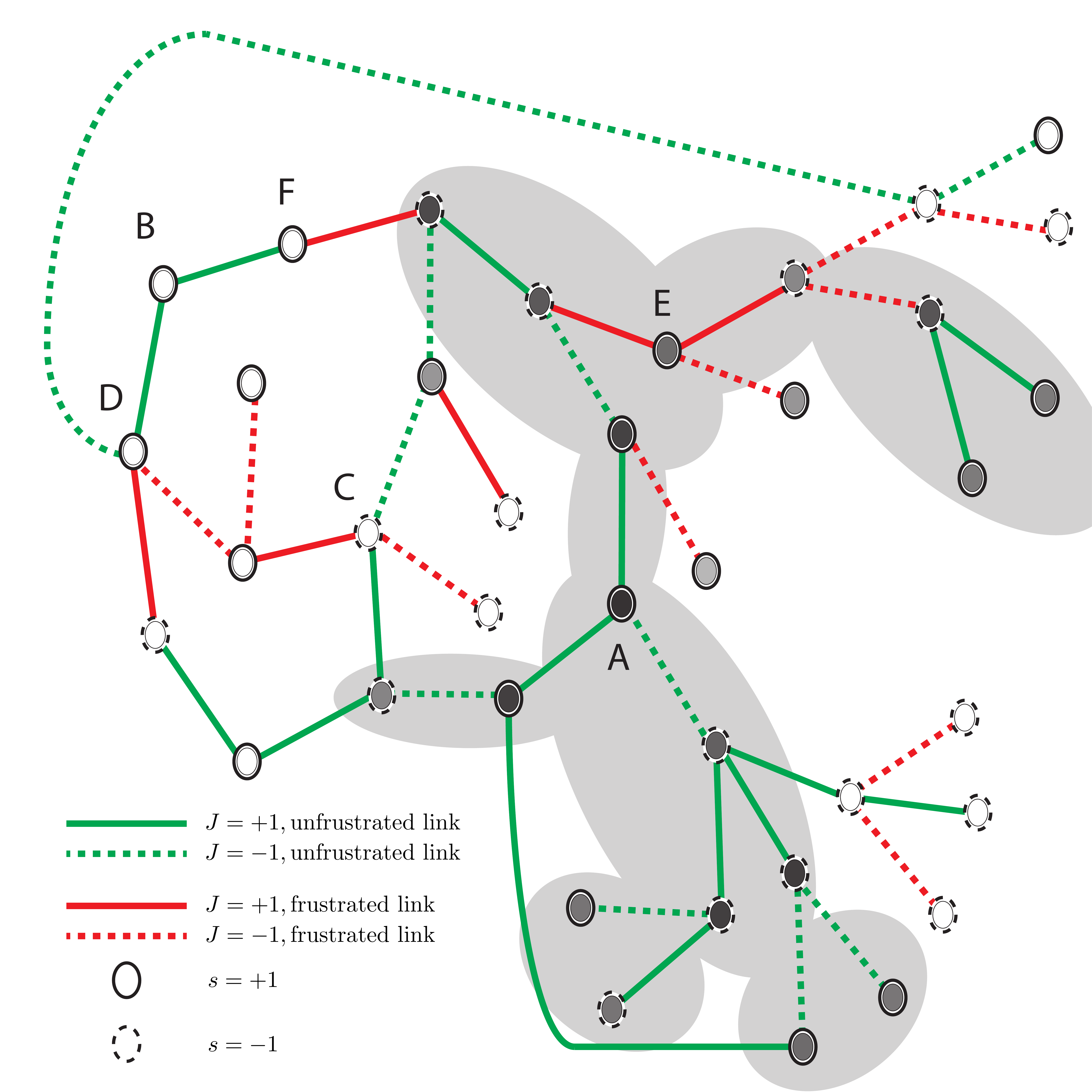}
	\caption{(Colour online) A toy model which comprises 38 spin variables. Unfrustrated links are marked in green and frustrated links in red. Positive $J$ and $s$ are denoted by solid lines and circles respectively, while negatives $J$ and $s$ are denoted by dashed lines and circles respectively. White nodes represents $p(\sigma_i\!=\!1)\!=\!0$ and black $p(\sigma_i\!=\!1)\!=\!1$; the grey region identifies nodes for which $p(\sigma_i) \!>\! 0.5$.}
	\label{fig_toymodel}
\end{figure}

\emph{Toy model:} 
To illustrate how the BP method identifies SSC nodes, we constructed a toy model comprising 38 spins as shown in Fig.~\ref{fig_toymodel}. In this case a specific configuration $\us$ is considered and the corresponding SSC-memberships marginals $p(\sigma_i),~\forall i$, are calculated.
Colours of the nodes represent {$p(\sigma_i)$} in a grey scale: white represents $p(\sigma_i\!=\!1)\!=\!0$ and black $p(\sigma_i\!=\!1)\!=\!1$.
The grey region identifies nodes for which $p(\sigma_i) \!>\! 0.5$.
We observe that spins which do not experience frustration (green regions) are more likely to belong to an SSC, but this is true only if their neighbours are also part of the SSC. For instance, while both nodes A and B do not experience frustration $p(\sigma_A) \!=\! 0.91$ while $p(\sigma_B) \!=\! 0$. It is instructive to investigate why the neighbours of B are not part of the SSC: nodes D and F receive conflicting messages from their neighbours and thus the local fields acting on them are 0, hence they cannot be part of an SSC and consequently node B is not part of an SSC {either}. The situation of node C is similar to that of node D. Interestingly, we observe that node E {\emph is} part of the SSC. This is counter-intuitive because this node is very frustrated, having all the links unsatisfied. At the same time, it can flip in order to satisfy all three links. Since two of its neighbours are part of the SSC and the three links can be satisfied by a single spin flip, node E is identified as being within the SSC, having $p(\sigma_E) \!=\! 0.66$. The last consideration is also clearly related to the notion of dynamical time scale separation; after one flip, spin E experience the same large absolute value of the local field, and will have all of its links satisfied. Thus, it should still be considered a slow spin.

\begin{figure*}
\includegraphics[width=180mm]{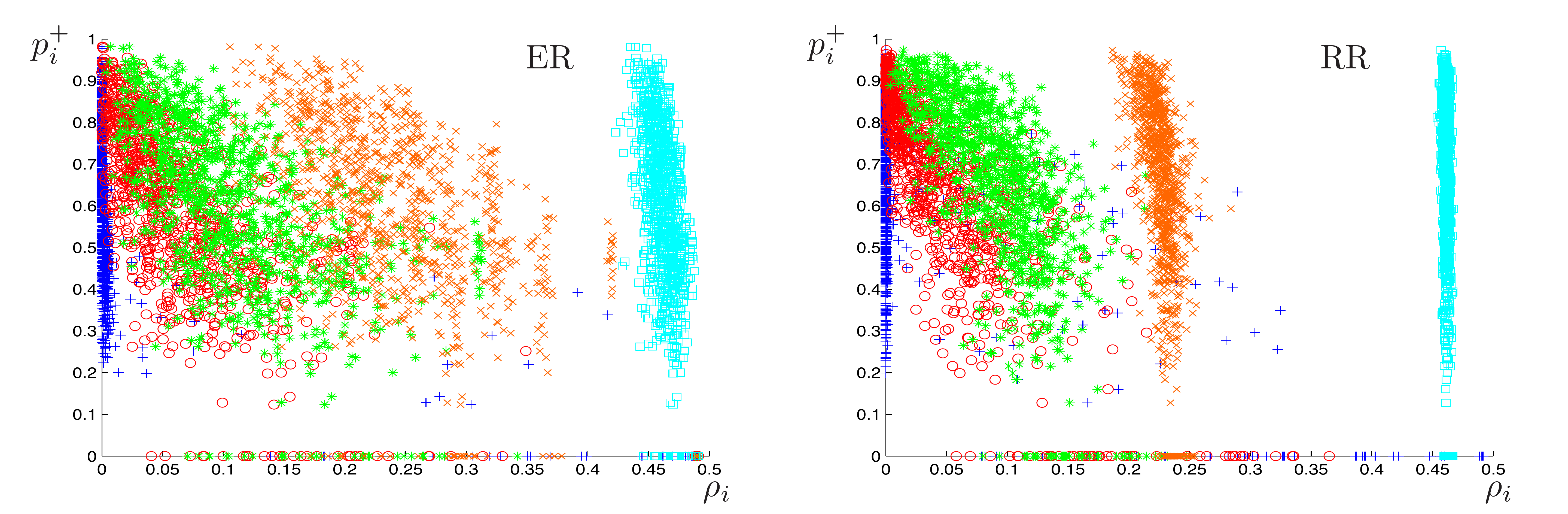}
\caption{Scatter plot of $p^+_i$ and $\rho_i$ for dynamics on a Erdos-Renyi graph of mean connectivity $6$ (left), random regular graph of connectivity $6$ (right). A configuration is sampled at $\beta\!=\!5$ after $t_w\!=\!10^4$ \textit{sweeps}. As discussed in the text, from this configuration we ran $1000$ simulations that we observed for $100$ Monte-Carlo Sweeps (MCS), that can been used to compute $\rho_i$. The symbols $\{ +, \tt{o}, *, \tt{x} \mbox{ and }\Box \}$ refer to dynamics at $\beta\!=\!3, 1, 0.7, 0.4 \mbox{ and }0.1$, respectively. Spin-glass transition inverse-temperatures are $\beta^{ER}_6 \!\sim\! 0.387$ and $\beta^{RR}_6 \!\sim\! 0.420$.}
\label{fig_Fig_c=6}
\end{figure*}

\emph{Results:} 
We consider spin glass models of $N\!=\!1000$ spins defined on Erdos-Renyi (ER) and random regular (RR) topologies and implement a sequential Glauber dynamics in the spin glass phase starting from a random configuration.
For (mean) connectivity equals to $c$, the critical inverse-temperatures for the spin-glass transition are known for both models~\cite{thouless1986spin} to be 
\begin{equation}
\beta^{ER}_c = \arctanh \left( \frac{1}{\sqrt{c}} \right) \:, \quad \beta^{RR}_c = \arctanh \left( \frac{1}{\sqrt{c-1}} \right)\:.
\end{equation}
After waiting $t_w$ dynamical \textit{sweeps}, where one \textit{sweep} is defined by updating the whole system, we sample a single configuration (the significance of $t_w$ is explained later).
In this configuration we study the SSC structure by running the BP algorithm with $\epsilon\!=\!0.1$, which allows to associate each spin with the corresponding SSC-membership probability $p^+_i \!=\! p(\sigma_i\!=\!1)$.
We conduct the following experiment: starting from the sampled configuration, we study $1000$ trajectories of the system at different inverse-temperatures for $T\!=\!100$ \textit{sweeps}; for each temperature we consider the number of flips per spin. 
We readily obtain the flipping rates $\rho_i$ by dividing the number of flips by $T\!=\!100$.
If the self-sustained structure contains information about the future evolution of the systems, flipping rates $\rho_i$ and SSC probabilities $p^+_i$ should be correlated, thus we plot both quantities for different temperatures. 

Figure~\ref{fig_Fig_c=6} shows $p^+_i$ and $\rho_i$ for ER and RR topologies of (mean) connectivity $6$, where $p^+_i$ are computed on the configuration sampled after $t_w\!=\!10^4$ \textit{sweeps} at $\beta\!=\!5$, and $\rho_i$ for dynamics starting from this configuration at $\beta\!=\!3,1,0.7,0.4,0.1$.
While for large $\beta$ there is a strong relation between spins with high SSC-membership  marginals $p^+_i$ and small flip-rate $\rho_i$, this relation disappears as $\beta$ decreases. This behaviour can be interpreted as follows: The SSC structure of the configuration sampled at very low temperatures after a long waiting time describes a metastable arrangement of spins with long-range correlations. At low temperatures the SSC spins remain close to the initial configuration, especially spins with $p^+_i \!\sim\! 1$, which flip less frequently than those with smaller $p^+_i$ values. 
The inverse spin-glass transition temperatures in these two cases are $\beta^{ER} \!\sim\! 0.387$ and $\beta^{RR} \!\sim\! 0.420$; this explains why in the ER case (left figure) the relation between $p^+_i$ and $\rho_i$ {remains} stable in a larger range of temperatures.
We also notice in both cases that the sampled configuration includes spins with $p^+_i\!=\!0$. Interestingly, these spins flip more frequently at high $\beta$ (for instance $\beta\!=\!3,5$) where the dynamics exhibits many slow variables and a few fast evolving ones. 
{These} results show that spins with a higher SSC probability tend to be dynamically more stable and vice versa. It further implies that SSC can be used as a tool to predict individual spin dynamics based on a single snapshot of equilibrium spin configuration.

\begin{figure*}
	\includegraphics[width=180mm]{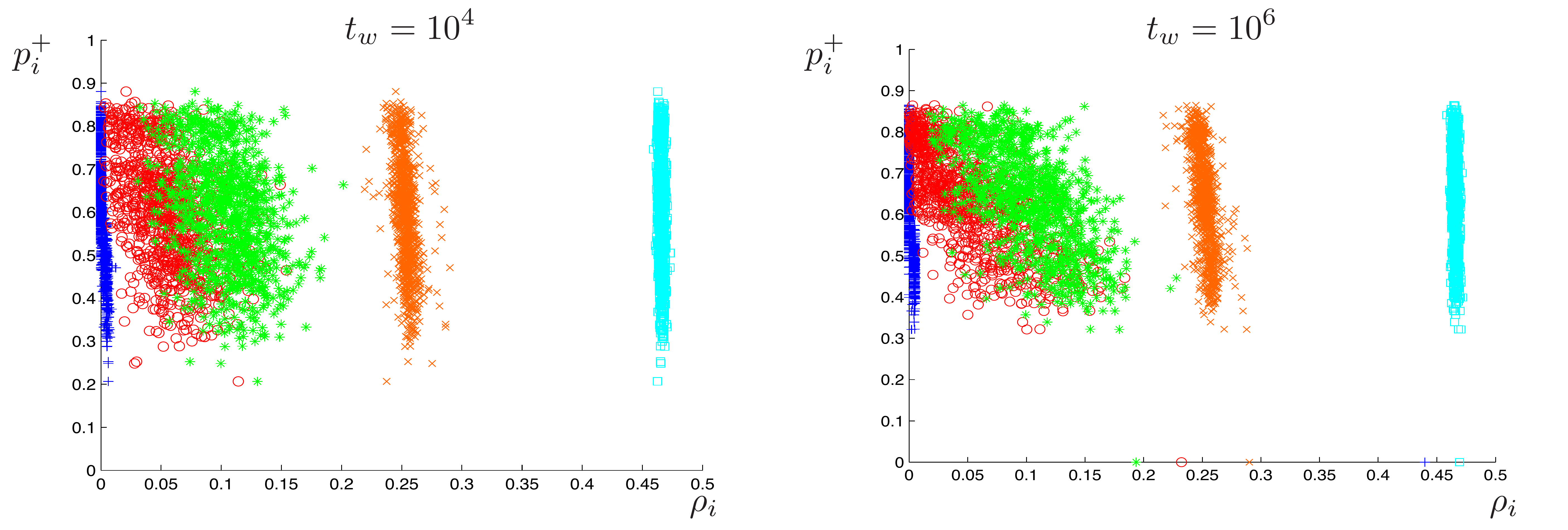}
	\caption{Scatter plot of $p^+_i$ and $\rho_i$ for dynamics on a RR graph of connectivity equal to $5$. A configuration is sampled at $\beta\!=\!5$ after $t_w\!=\!10^4$ \textit{sweeps} (left) and after $t_w\!=\!10^6$ \textit{sweeps} (right). As discussed in the text, from this configuration we ran $1000$ simulations that we observed for $100$ {MCS}, that can been used to compute $\rho_i$. The symbols $\{ +, \tt{o}, *, \tt{x} \mbox{ and }\Box \}$ refer to dynamics at $\beta\!=\!3, 1, 0.7, 0.4 \mbox{ and }0.1$, respectively. Inverse spin glass temperature in this case is $\beta^{RR}_5 \sim 0.464$.}
	\label{fig_Fig_c=5}
\end{figure*}

To further examine the {predictive} power of SSC probabilities, we compute them on spin configurations sampled after two different waiting times $t_w\!=\!10^4$ and $t_w\!=\!10^6$ \textit{sweeps} for RR topologies of connectivity $5$.  { Figure~\ref{fig_Fig_c=5} shows that the behavior described above is more emphasized when longer waiting times are used; i.e., data points (of $p_i^+$ vs. $\rho_i$ ) at low temperatures have steeper and more well defined slopes. It also implies that the snapshots of SSC structures taken after a long time are more stable and hence have a higher predictive power of stable spins. As the rationale of SSC can be extended to systems in other areas, these results suggest that the SSC methodology can be used to devise generic tools for predicting the dynamical properties of variables in a wide range of applications based on limited static information.
We also notice that the time scale separation between fast and slow spins at low temperatures disappears in this case. Consistently, our algorithm, does not find spins where $p^+_i\!=\!0$.}

In order to investigate further this effect, we extended the analysis to RR topologies of {similar} connectivities studying the cases of $c=3,4$. We find that time scale separation at low temperatures {appear} only for even connectives in RR graphs. This is discussed further in the SM, together with more results on the dynamics {of systems} on ER graphs.

\emph{Summary:} We developed a theoretical framework linking dynamical and equilibrium properties of spin-glass models {on sparse graphs} based on the concept of SSC, which can be viewed as regions of interdependent mutually-stabilizing spins. We show that the SSC structure of a given sampled configuration predicts {the} dynamical properties of the system, such as the microscopic time-scale separation in certain systems (low temperature dynamics on RR topologies with even connectivities or ER graphs) and regions of slow-evolving variables in spin systems. 
We provide a new microscopic perspective on the low temperature dynamics of spin glass systems with the potential of developing new algorithmic optimization tools for hard computational problems through the destabilization of SSC.  {Furthermore, our results show that the SSC paradigm} can be used as a tool to predict individual spin dynamics based on a single snapshot of equilibrium spin configuration, which can be extended to applications in other areas.

\emph{Acknowledgements} This work is supported by The Leverhulme Trust (RPG-2013-48) and the Research Grants Council of Hong Kong (CHY, 28300215, 18304316). We thank F. Ricci-Tersenghi for interesting discussions.

\bibliography{apssamp}% Produces the bibliography via BibTeX.
\bibliographystyle{unsrt}

\onecolumngrid
\appendix

\section*{Supplementary Material}

\subsection{Factor graph and BP equations}
\begin{figure}[ht]
	\includegraphics[width=85mm]{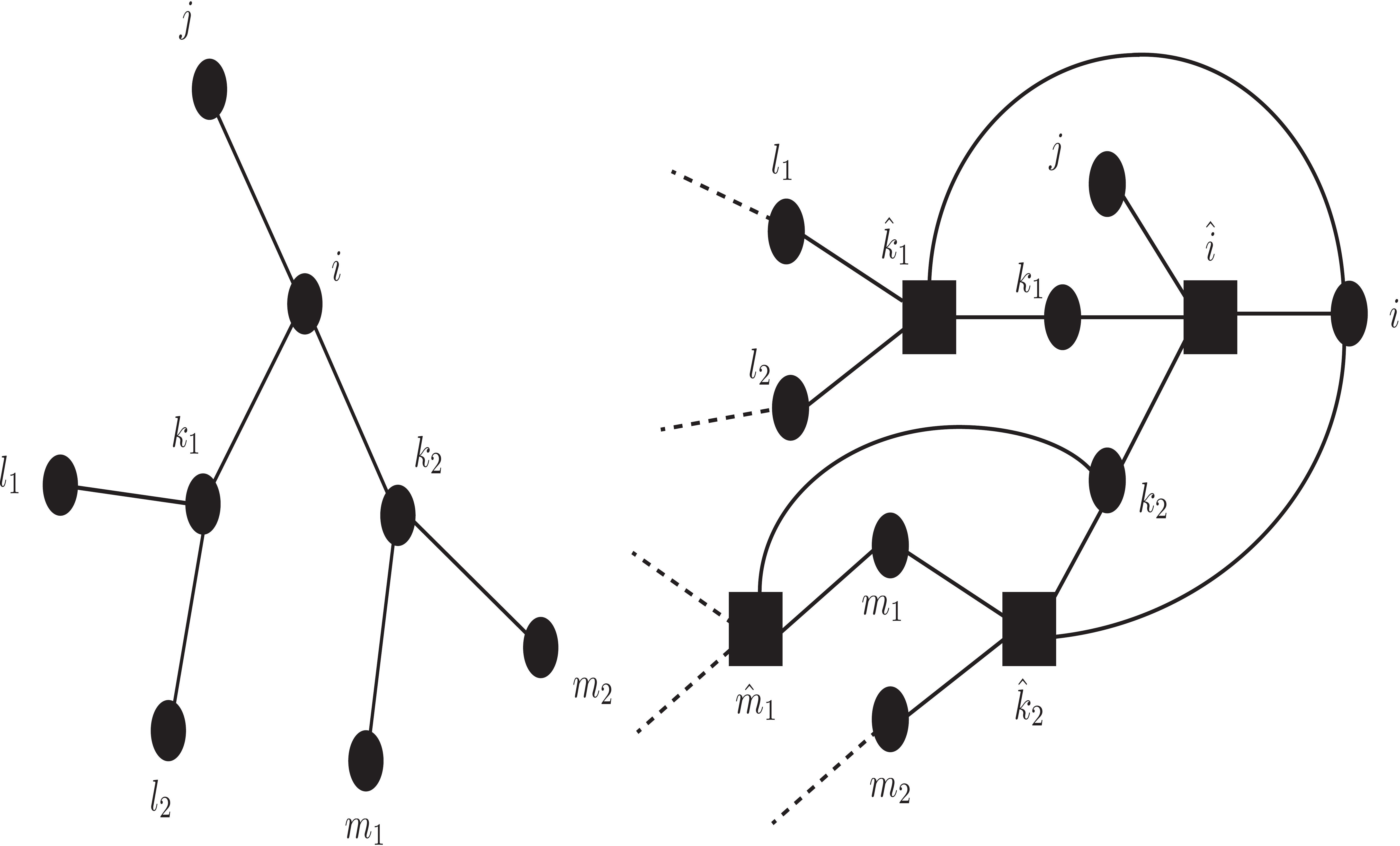}% Here is how to import EPS art
	\caption{\label{fig:sfig1} The original graph topology (left) specified by the tree $\mathcal{G}$ and corresponding factor graph for the SSC problem (right), which clearly contains many small loops.}
	\label{fig:graphandSSCFG}
\end{figure}

This section explains the need for a super-factor graph of super-nodes that comprise variable pairs due to the emergence of many loops if single-variable factor graph is introduced. Let un consider the graph in left side of Fig.~\ref{fig:graphandSSCFG}. On this graph, we would like to compute the marginals for $\sigma$'s values. To this aim, we introduce the factor graph on the right side of  Fig.~\ref{fig:graphandSSCFG}. Typically, the message factor $\hat{i}$ sends to node $i$ can be written as
\be
\nu_{\hi \ra i} ( \sigma_i ) \propto \sum_{\ux_{\partial \hi \bsl i}} \psi_{\hi} (\ux_{\partial \hi}) \prod_{k \in \partial \hi \bsl i } \eta_{k \ra \hi} (\sigma_k)\:, 
\ee
while the message node $k$ sends to factor $\hi$ becomes 
\be
\eta_{k \ra \hi} (\sigma_k) \propto \prod_{\hk \in \partial k \bsl i } \nu_{\hk \ra k} (\sigma_k)\:.
\ee
This factor graph contains small loops due to the dependence of variables on the neighbors of their neighbors. 
In other words, the state of variable $\sigma_i$, depends on the state of all of its neighbors, whose state depend on $\sigma_i$ itself.

This is a notorious problem for implementing BP and inferring the related  marginals. To overcome this problem we introduce a modified factor graph with super-variables formed by variable pairs as explained in the main text.
%where there are not small loops, since its structure is the same as that of the original graph. We can also observe that this super-graph can be created by considering a super-factor for each node, and a super-variable for each link of the original graph.
%

\subsection{Dynamics on RR graphs with even and odd connectivities}

As {described} in the main text, the time-scale separation between slow and fast spins observed at low temperatures for (mean) connectivity $6$ in Fig. ~\ref{fig_Fig_c=6} does not appear in the RR case with connectivity $5$, see Fig. ~\ref{fig_Fig_c=5}, where (almost) all  spins are blocked or flip at most once at $\beta\!=5\!$. Consistently, we do not find spins with SSC probability {$p_i^+ \!=\! 0$}. We attribute this behavior to the strong residual field in odd degree connections.
To support this conjecture we consider RR graphs of connectivities $3$ and $4$, observing the expected behavior as shown in Fig.~\ref{fig_Fig_c=3,4}. 

\begin{figure*}[ht]
  \includegraphics[width=180mm]{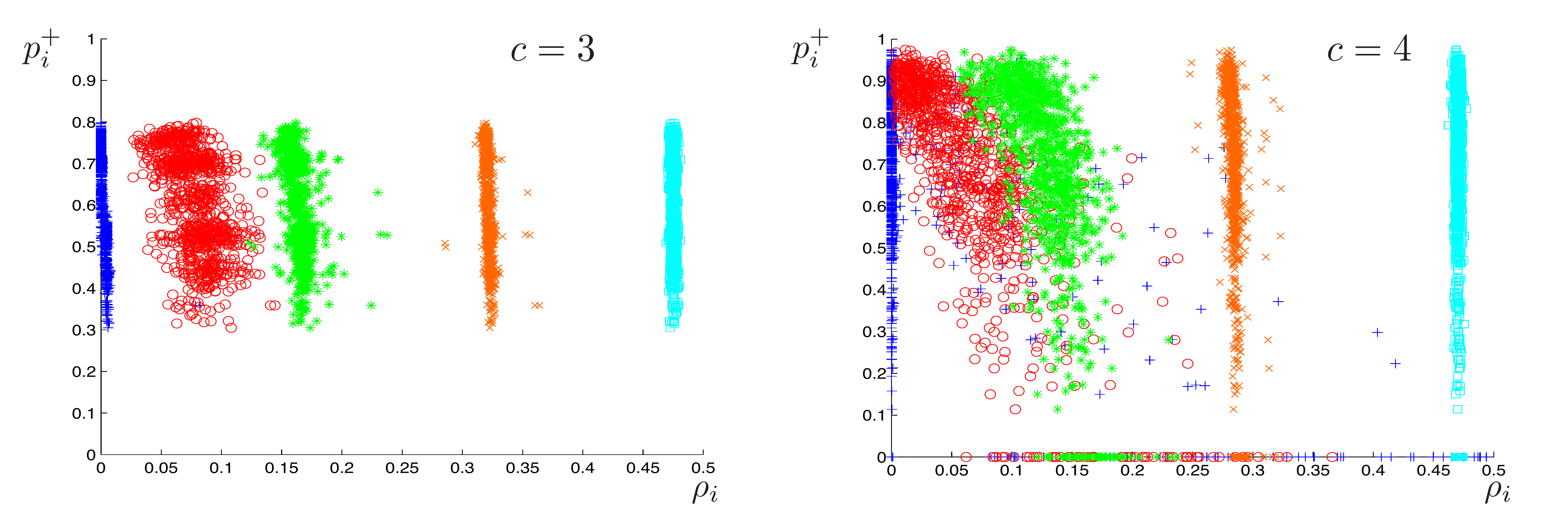}
\caption{Scatter plot of $p^+_i$ and $\rho_i$ for the dynamics on an RR graph of connectivities $3$ (left) and $4$ (right).  A configuration is sampled at $\beta\!=\!5$ after $t_w\!=\!10^4$ \textit{sweeps}. As discussed in the text, from this configuration we ran $1000$ simulations that we observed for $100$ {MCS}, that can been used to compute $\rho_i$. The symbols $\{ +, \tt{o}, *, \tt{x} \mbox{ and }\Box \}$ refer to dynamics at $\beta\!=\!3, 1, 0.7, 0.4 \mbox{ and }0.1$, respectively. Spin-glass transition inverse-temperatures are $\beta^{RR}_3 \!\sim\! 0.615$ and $\beta^{RR}_4 \!\sim\! 0.524$.}
\label{fig_Fig_c=3,4}
\end{figure*}

\subsection{Dynamics on ER topologies}

\begin{figure*}
	\includegraphics[width=180mm]{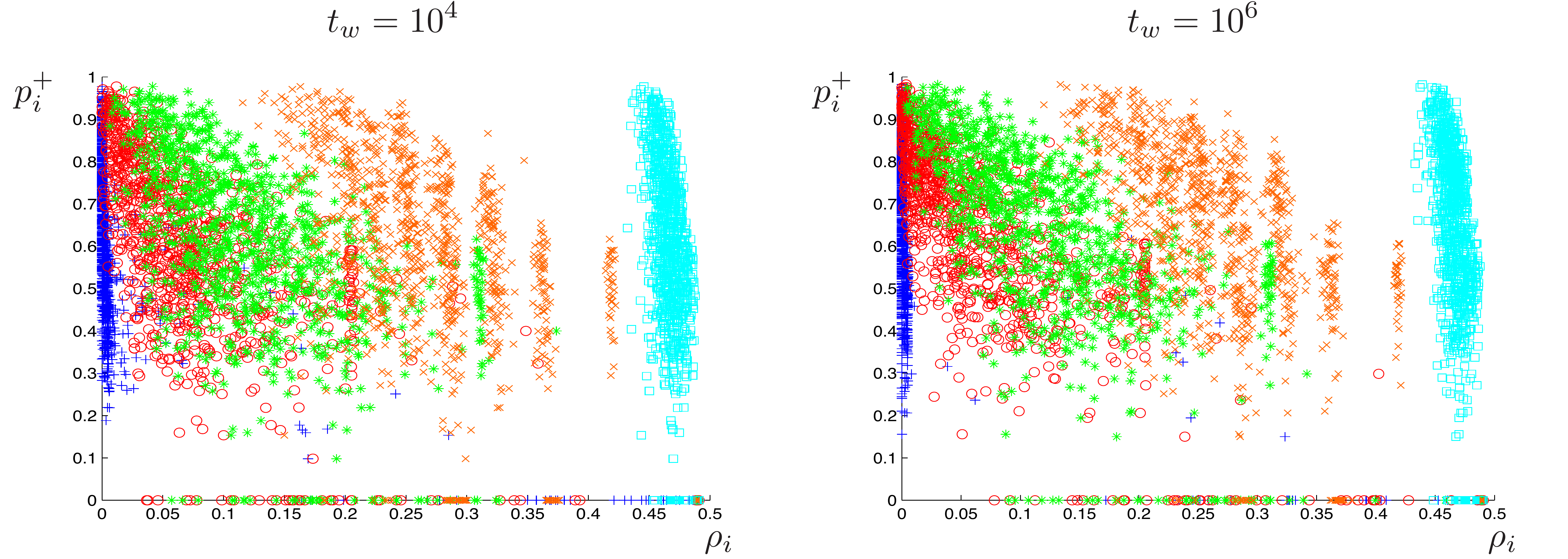}
	\caption{Scatter plot of $p^+_i$ and $\rho_i$ for dynamics on a ER graph of mean connectivity equal to $5$. A configuration is sampled at $\beta\!=\!5$ after $t_w\!=\!10^4$ \textit{sweeps}  (left) and after $t_w\!=\!10^6$ \textit{sweeps} (right) . As discussed in the text, from this configuration we ran $1000$ simulations that we observed for $100$ {MCS}, that can been used to compute $\rho_i$. The symbols $\{ +, \tt{o}, *, \tt{x} \mbox{ and }\Box \}$ refer to dynamics at $\beta\!=\!3, 1, 0.7, 0.4 \mbox{ and }0.1$, respectively. Inverse spin glass temperature in this case is $\beta^{ER}_5 \sim 0.420$.}
	\label{fig_FigER_c=5}
\end{figure*}

\begin{figure*}
	\includegraphics[width=180mm]{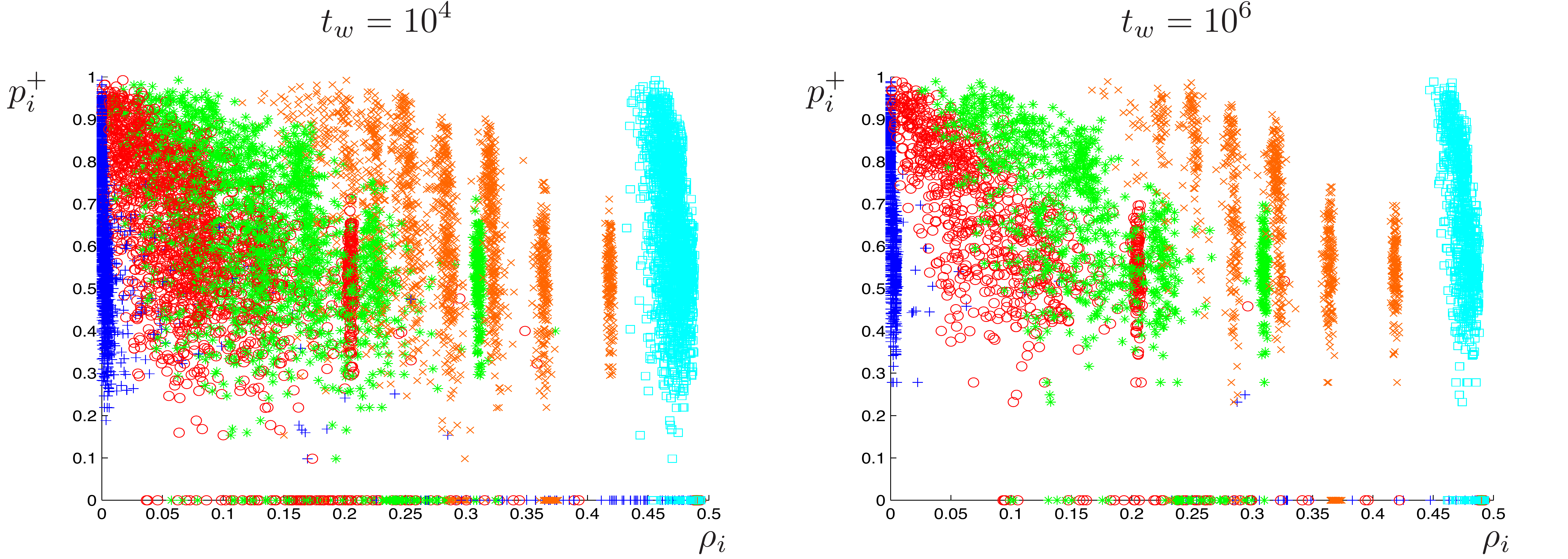}
	\caption{Scatter plot of $p^+_i$ and $\rho_i$ for dynamics on a ER graph of mean connectivity equal to $3$. A configuration is sampled at $\beta\!=\!5$ after $t_w\!=\!10^4$ \textit{sweeps}  (left) and after $t_w\!=\!10^6$ \textit{sweeps} (right). As discussed in the text, from this configuration we ran $1000$ simulations that we observed for $100$ {MCS}, that can been used to compute $\rho_i$. The symbols $\{ +, \tt{o}, *, \tt{x} \mbox{ and }\Box \}$ refer to dynamics at $\beta\!=\!3, 1, 0.7, 0.4 \mbox{ and }0.1$, respectively. Inverse spin glass temperature in this case is $\beta^{ER}_3 \sim 0.523$.}
	\label{fig_FigER_c=3}
\end{figure*}

In the ER case, the dynamics at $\beta \!\sim\! \beta^{ER}$ exhibits spins that flip at specific frequencies. This effect is not visible in the RR case and is due to {its degree heterogeneity} as can be observed in Figs.~\ref{fig_FigER_c=5}-\ref{fig_FigER_c=3}.

\end{document}